# Particle Physics and Astrophysics

A whitepaper in response to a call to the Astronomy and Astrophysics Community from the Committee on Astro2010 for State of the Profession Position Papers

March 2009

L. Buckley-Geer (a), P. Burchat (b), D. Burke (c), E. Cheu (d), D. Cinabro (e), K. Honscheid (f), S. Kahn (b,c), D. Kirkby (g), B. Meadows (h), R. Schindler (c), I. Shipsey (i), J. Thaler (j), W. Wester (a)

(a) Fermi National Accelerator Laboratory, (b) Stanford University, (c) SLAC National Accelerator Laboratory, (d) University of Arizona, (e) Wayne State University, (f) The Ohio State University, (g) University of California, Irvine, (h) University of Cincinnati, (i) Purdue University, (j) University of Illinois at Urbana-Champaign.

**Abstract**

We discuss some of the key science questions that are bringing particle physicists and astrophysicists together, and comment on some of the cultural and funding issues that have arisen as these two communities become increasingly intertwined.

**Contents**

**I. Particle Physics and Astrophysics**

The cosmos is a laboratory to probe the fundamental laws of physics in unique ways that are complementary to accelerator-based particle physics experiments. Observing the cosmos has provided strong evidence for physics not contained in the Standard Model of particle physics: non-zero neutrino masses, non-baryonic dark matter, dark energy and primordial inflation -- the latter likely tied to the unification of the disparate forces in the Universe near the Planck scale. Together with ordinary matter, these elements provide a consistent accounting of the constituents of the Universe. The interactions of these constituents have profoundly shaped the structure of the Universe over eons of cosmic time, and still do so today. The necessary, and yet puzzling, connections between the inner space of quantum reality and outer space of cosmic reality enable the discovery of new particle physics through astrophysical observations. It is for this reason that particle physicists have been inexorably drawn to astronomy in recent times.

A combination of observatories in space and on the ground, accelerator-based experiments and non-accelerator experiments are necessary to address the fundamental physics questions raised by the new paradigm:

• What is dark matter?
• Why is the expansion of the Universe accelerating?
• What are the neutrino masses and what is their impact on cosmic evolution?
• Was an epoch of primordial inflation responsible for the origin of large-scale structure? Did it leave observable imprints that can shed light on the unification of the fundamental particles and forces at energies that far exceed those accessible to terrestrial accelerators?

**Dark Matter**
Dark matter (DM) lies at the confluence of particle physics and astrophysics, and is of profound importance to both fields. DM's crucial role in the evolution of the Universe and the development of structure ensure its central position in the study of cosmology. Simultaneously, cosmological constraints on the cosmic baryon density imply that most of the DM is non-baryonic; therefore DM requires physics beyond the Standard Model, making its elucidation a crucial goal of particle physics. The search for DM particles involves three main independent and complementary approaches: indirect searches, which seek evidence for DM annihilations occurring near the center of our galaxy and in other astrophysical bodies; direct searches, which seek evidence for DM particles from the halo of our galaxy interacting with terrestrial detectors; and accelerator-based searches, which seek to create DM particles and probe their quantum mechanical properties in laboratory experiments. The exciting prospect of bringing these three approaches to convergence in a single understanding of DM and its properties is within reach of existing and near-term experiments; this would be one of the great scientific achievements of the second decade of the 21$^{st}$ century.

Properties of DM can be gleaned by studying the evolution of structures, such as galaxies and galaxy clusters, in the Universe. Models of structure formation indicate that the bulk of the DM has likely been non-relativistic over most of cosmic history; it is therefore known as cold dark matter (CDM). While the mass density of DM in the Universe is becoming known to high precision, the identity of the hypothetical CDM particle remains a complete mystery. No particle contained within the laws of physics as we know them today has the properties to be CDM. However, CDM particles do emerge naturally in a variety of well-motivated theories of physics beyond the Standard Model. The two most compelling candidates are axions and weakly interacting massive particles, or WIMPs.

Axions are elementary particles that arise in theories that explain why we have not detected large asymmetries between matter and anti-matter in the strong interactions of the Standard Model. Axions are expected to be less massive and more weakly interacting than neutrinos. The axion can be detected by its transformation into a photon in the presence of a magnetic field. As an example, the Axion Dark Matter experiment (ADMX) recently ruled out an axion with mass of about $10^{-6}$ eV[i]. However, experiments are not yet sufficiently sensitive to fully probe the mass regions expected if the DM particle is an axion.

WIMPs are elementary particles that interact with ordinary matter with a cross section of similar magnitude to the weak interaction cross section of the Standard Model. The success of the electroweak theory suggests that unification can be taken one stage further and the strong force unified with the electroweak force. A leading unification scheme invokes supersymmetry (SUSY)[ii], which predicts that for every fundamental particle known, there is a SUSY partner. The neutralino, the lightest stable SUSY particle in a wide class of SUSY models, with an expected mass between 100 GeV and 1 TeV, is a WIMP and a prime DM candidate. Indirect and direct searches for WIMPs are becoming increasingly sensitive. Among the direct searches, CDMS, for example, is probing the upper range of cross sections that can be accommodated in the simplest SUSY extensions of the Standard Model[iii]. One of the most exciting possibilities enabled by the Large Hadron Collider (LHC) is the production and detection of candidate DM particles[iv]. If WIMPs are discovered at the LHC, measurements at the LHC, and eventually at a lepton collider as well, will allow the deduction of the relic density of WIMPS. This probes the composition of the cosmic DM sector, which could provide further constraints on models of the formation of galaxies and on the evolution of the Universe[v].

**Dark Energy**
The accelerating expansion of the Universe was Science Magazine's discovery of the year in 1998. Subsequent observations have independently confirmed and amplified this remarkable finding. Two possibilities could account for the accelerating expansion: either three quarters of the energy density of the Universe is in a new form called Dark Energy (DE), or general relativity breaks down on cosmological scales and must be replaced with a new theory of gravity. Either way there are profound implications for our understanding of the cosmos and of the fundamental laws of physics.

DE could be the energy of the vacuum, equivalent to Einstein's cosmological constant. Although sometimes considered the simplest model for DE, conventional particle physics theory predicts that the vacuum energy density should be many orders of magnitude larger than the value that would account for the present acceleration[vi]. This mismatch is a profound challenge to our understanding of quantum reality. Alternatively, DE could signal the existence of a new, ultra-light particle not in the Standard Model, an idea known as quintessence.

While the nature of DE is unknown, a well-defined set of first questions has emerged: Is DE the cosmological constant? Is it energy or gravity? Do its properties evolve over time? DE experiments address these questions by studying the impact of DE on both the history of the cosmic expansion rate and the growth rate of large-scale structure. In general relativity, the expansion rate, and the properties of DM, determine the rate at which structure forms; departures from Einstein's gravity may thus be tested by comparing the expansion rate history with the history of structure growth.

**Neutrinos**
Neutrinos are of great interest to both particle physicists and astronomers, not only as objects of study, but also as probes of other phenomena. Studies in one field often lead to developments in the other. A striking example of this is the discovery, using neutrinos produced by the Sun and by cosmic rays, that neutrinos are not massless. This was the first direct evidence for physics beyond the Standard Model of particle physics.

Today, studies of neutrinos in particle physics focus on mass and mixing. These properties have potentially significant effects on the cosmological large-scale structure[vii] and matter-antimatter

asymmetry[viii]. Conversely, cosmological studies put significant constraints on the neutrino mass[ix] and test some theories beyond the Standard Model of particle physics[x].

The new large neutrino "telescopes" being planned (for example, IceCube) will open a new window to astrophysical processes. Cosmic ray neutrinos are promising probes of supernovae and SN remnants[xi], as well as gamma ray bursts and AGN[xii]. They may also aid in dark matter (WIMP annihilation) searches[xiii].

**The Planck Scale**
Significant effort in theoretical physics is devoted to identifying and understanding the sensitivity of astrophysics probes of models of inflation. One probe is the polarization of the Cosmic Microwave Background (CMB). The rapid expansion of the Universe during inflation is thought to have generated gravitational waves. These waves are difficult to detect directly. However, when the CMB photons were last scattering, gravity waves induced motions in the photon-baryon fluid, imprinting patterns in the polarization of the CMB. Non-Gaussianity in large-scale structure (LSS) is another possible probe of models of inflation. Future, more sensitive studies of the CMB and LSS may shed light on the unification of the fundamental particles and forces at energies that far exceed those accessible to terrestrial accelerators.

## II. Recommendations from previous surveys and panels relevant to participation of particle physicists in astrophysics and cosmology

In the 2001 decadal survey (Astronomy and Astrophysics in the New Millennium), the Survey Committee included a recommendation for the Department of Energy in the chapter on Policy for Astronomy and Astrophysics:

> *"As the size and complexity of astronomy and astrophysics projects increase, funding patterns are changing in ways that challenge traditional agency boundaries and funding patterns, and interagency collaborations are frequently advantageous. The committee commends DOE for supporting astrophysical research and recommends that DOE develop a strategic plan for astrophysics to ensure a vigorous, coherent research program and to facilitate cooperation with other agencies."*

Strategic planning in particle physics and particle astrophysics in the DOE and NSF depends on the advice of the High Energy Physics Advisory Panel (HEPAP) and the Astronomy and Astrophysics Advisory Committee (AAAC), and reports of the National Academy of Sciences and other scientific organizations such as the American Physical Society. HEPAP advises the federal government on the national research program in experimental and theoretical particle physics research, and reports to both the Associate Director of the DOE Office of High Energy Physics in the Office of Science, and the Assistant Director of the NSF Mathematical & Physical Sciences Directorate. HEPAP regularly appoints subpanels that are charged to tackle specific issues facing the field. Subpanels are sometimes charged jointly by two advisory groups; a recent example is the 2007 Dark Matter Scientific Assessment Group, which was charged by HEPAP and AAAC.

In June 2006, a HEPAP subpanel was formed to review the DOE and NSF high energy physics

University Grants Program. As part of this study, the panel conducted a survey in January 2007 of 407 DOE and NSF investigators in experimental and theoretical particle physics; 268 (66%) investigators responded. The survey included questions about the investigators' current distribution of research effort and anticipated research effort in 2012. The results demonstrated a significant anticipated shift towards astrophysics and cosmology[xiv]. By 2012, half of the FTE effort in experimental particle physics is expected to be devoted to the Large Hadron Collider, while the second-largest effort in 2012 is expected to be in astrophysics and cosmology, followed by neutrino physics, linear collider development, underground physics, heavy quark physics, and Tevatron physics, in that order. For particle theorists, the largest research area, both in 2007 and anticipated in 2012, is particle phenomenology, followed by astrophysics and cosmology, with significantly more effort expected in these two areas than in the other subfields (string theory, field theory, model building and QCD/lattice QCD).

In 2006, a National Research Council Committee on Elementary Particle Physics in the 21$^{st}$ Century issued a report in which they recommended that DOE and NSF work together to achieve four objectives, in priority order: fully exploit the opportunities afforded by the LHC, plan a program to become the world-leading center for R&D for a linear collider and bring it to the US, and *"expand the program in particle astrophysics and pursue an internationally coordinated, staged program in neutrino physics"*.

A HEPAP subpanel called the Particle Physics Project Prioritization Panel (P5) was first constituted in 2002 and has regularly conducted long-range planning exercises. In its most recent report (US Particle Physics: Scientific Opportunities, A Strategic Plan for the Next Ten Years), issued in May 2008, P5 focused on three frontiers of particle physics: the energy, intensity and cosmic frontiers. In the executive summary, P5 included the following recommendations regarding the cosmic frontier:

- *The panel recommends support for the study of dark matter and dark energy as an integral part of the US particle physics program.*
- *The panel recommends that DOE support the space-based Joint Dark Energy Mission, in collaboration with NASA, at an appropriate level negotiated with NASA.*
- *The panel recommends DOE support for the ground-based Large Synoptic Survey Telescope program in coordination with NSF at a level that depends on the overall program budget.*
- *The panel further recommends joint NSF and DOE support for direct dark matter search experiments.*
- *The panel recommends limited R&D funding for other particle astrophysics projects and recommends establishing a Particle Astrophysics Science Advisory Group.*

Very recently, the DOE and NSF responded to these recommendations by charging a new HEPAP subpanel, the Particle Astrophysics Scientific Assessment Group (PASAG). The charge was presented to HEPAP by Dennis Kovar, Associate Director of the DOE Office of Science for High Energy Physics, on February 25, 2009. According to the materials presented to HEPAP, the scientific scope of the review "should be limited to opportunities that will advance our understanding of the fundamental properties of particles and forces using observations of phenomena from astrophysical sources." Scientific areas considered within the scope of the study include "exploring the particle nature of dark matter, understanding the fundamental properties of dark energy, and measuring the properties of astrophysically generated particles (including cosmic rays, gamma rays, and neutrinos)." It was recognized that some of the research areas identified are within the scope of Astro2010; "appropriate sharing of information should be explored." PASAG is asked to provide recommendations on priorities under four scenarios for funding profiles, and to provide preliminary comments by July 1, 2009 and a final report by August 15, 2009.

**III. A New Blended Culture**

The number of particle physicists taking active roles in astrophysics has increased significantly. The Sloan Digital Sky Survey (SDSS) had about 10% particle physics membership, while the percentages on DES and LSST are closer to 50%. We think that this cross-disciplinary interaction is a healthy development, benefiting both communities. However, the inevitable stresses[xv] created by this "immigration" cannot be ignored. While it is true that there is a significant cultural difference between particle physics and astronomy, we think that this difference is being bridged with a stronger, blended culture emerging.

We perceive two significant issues: the evolution toward large collaborations, and the relationship between instrument builders and observers. Astronomy has relied (and still does rely, in many areas) on facilities constructed by instrument builders who do not necessarily participate in specific science proposals. Observers propose the creation of specific, targeted data sets and then perform the data analysis. They are the only ones who author the resulting papers. Particle physics does not operate this way. Instead, teams of scientists build large multipurpose detectors and then participate in the scientific data analysis. Typically, every collaboration member appears on every science paper, whether or not they participated in the particular data analysis.

Both the Dark Energy Survey (DES) and the Large Synoptic Survey Telescope (LSST) collaborations consist of large numbers of astronomers and particle physicists. The particle physicists often view these projects as similar to large particle physics projects (*e.g.*, BABAR or CDF) and anticipate that the collaborations will operate similarly. There are differing expectations, particularly regarding membership, authorship, and data rights. Both DES and LSST recognized the need to develop common expectations while their respective collaborations were still nascent. SDSS has served as a template, providing many examples of successful practices that DES and LSST have found helpful to adapt and adopt.

Multicultural collaborations bring many benefits. Particle physicists have a wealth of experience creating, organizing, and managing large international scientific collaborations. These collaborations are essential for the construction and commissioning of state of the art particle detectors, data acquisition systems, and distributed computing. The collaborations facilitate the creation of the large organized analysis teams required to expeditiously extract particle physics from very large data sets. Particle physicists bring technical infrastructure to astronomical projects; for example, silicon detector fabrication facilities at Fermilab and Brookhaven, computing infrastructure, and analysis tools such as the ROOT software package from CERN, which may be well-suited to the large data sets. Astronomers have an intimate knowledge of astronomical apparatus, and essential and critical skills in the analysis of astronomical data. Some examples include telescope design, spectroscopy, photometry, image reduction and analysis, and follow-up observing campaigns. Crucially, astronomers are well-versed in the subtle issues that can bite the naïve data analyst. The combined efforts of astronomers and particle physicists will broaden the scientific scope, and increase the scientific productivity, of these newly formed collaborations.

## IV. Funding particle physicists in astronomy

Approximately two-thirds of University groups working in experimental particle physics are funded through the Department of Energy (DOE) and the remainder through the National Science Foundation (NSF). In addition, DOE supports particle physicists at several national laboratories. Experimental particle physicists generally find that NSF research priorities are investigator driven, in contrast to DOE HEP, where the funding decisions appear to be more project focused. In a few cases, the choice of funding agency for University groups or individual investigators is determined by the support of a facility (*e.g.*, the NSF-sponsored group at Cornell working at the CESR storage ring), but in many cases it is largely 'historical' (*e.g.*, a new investigator receiving an NSF CAREER Award versus a DOE OJI Award). In this section, we discuss funding issues for particle physicists who wish to pursue research in astrophysics or cosmology through the NSF or DOE.

**National Science Foundation**
Experimental particle physics is supported by the NSF through two programs in the Physics division (PHY): Elementary Particle Physics (EPP) and Particle and Nuclear Astrophysics (PNA). Particle theorists are funded through the Theoretical Physics program in PHY. The PNA program was established in FY2000. Within PHY, the PNA program is the most relevant to the Astro2010 decadal survey. The synopsis of the PNA program[xvi] describes the current portfolio:

> *"Currently supported activities are: ultra high energy cosmic-ray and gamma-ray studies, the study of gamma-ray bursts, solar, underground and reactor neutrino physics, searches for the direct detection of Dark Matter and searches for neutrino-less double beta decay. Funding is also provided for accelerator-based nuclear astrophysics studies of stellar processes, nucleosynthesis, and processes related to cosmology and the early universe. Currently, the Deep Underground Science and Engineering Laboratory (DUSEL) program is funding Research and Development studies for experiments that need to be located in an underground laboratory to reduce backgrounds."*

There are currently 115 active PNA awards, including 36 for R&D for underground experiments (mostly DUSEL); roughly 15% of PNA awards are funded jointly with another NSF program. The funding level for PNA has been roughly $15M to $16M for the past four years, not including DUSEL R&D. Over half this funding supports searches for dark matter particles and studies of cosmic rays, high energy gamma rays and ultra high energy neutrinos. A small fraction of the funding supports physicists whose programs are at least partially directed towards observational cosmology and/or dark energy. One issue that arises as particle physicists seek support to pursue research in astrophysics or cosmology, motivated by questions related to particle physics, is where to seek funding. We advocate that particle physicists traditionally funded through NSF be encouraged to apply to the PNA program[1]. There is already a strong record of success in supporting efforts in astrophysics through PNA, but the results have been mixed for those motivated by cosmology and investigations of dark energy. In particular, particle physicists are presented with a severe challenge if they write a proposal with the expectation that it will be reviewed by the PNA community, but the proposal is passed to a different program, possibly outside the Physics Division (*e.g.*, AST). Because of the different expectations in the particle physics and astronomy communities for early support of individual investigators for R&D, planning and construction related to large projects, a proposal written with a particle-physics reviewer in mind is understandably unlikely to be highly rated among AST reviewers, when compared to more traditional AST proposals.

We feel that it is not feasible for particle physicists to write proposals that simultaneously satisfy the cultural expectations of the particle physics and astronomy communities. We therefore suggest that proposals submitted by particle physicists to pursue research related to cosmology and dark energy be reviewed within PNA. The community of particle physicists with expertise in this area is now broad and deep enough to support the PNA proposal review process, which is based on several individual "ad hoc" reviews by experts, followed by a final review by a broader panel. The traditional priorities of a typical PNA review panel are an excellent match to the needs of the very large projects being designed and constructed to tackle issues related to astrophysics, cosmology and dark energy that motivate particle physicists. The PNA review of and support for DUSEL R&D and underground physics present a relevant example.

**Department of Energy**
The Office of High Energy Physics (OHEP) within the Office of Science at DOE is the dominant source of funding for particle physics research in the United States. OHEP supports this field through the maintenance of facilities and funding of research at five national laboratories (Fermilab, SLAC, Brookhaven, Argonne, and LBNL), and through the University Grants Program, which issues "block grants" for the support of particle physics groups at individual universities. The university grants are three-year renewable, and are generally geared to coordinated efforts in a limited number of "key task areas" associated with participation in particular experiments (*e.g.,* BABAR, CDF, ATLAS), with multiple faculty members aligned with each task. In general, OHEP provides strong guidance to its university groups – they view this program as an integral piece of the national effort in particle physics research, and strive to ensure that the critical needs of the large experiments they support are being met using both laboratory and university resources.

As indicated above, the "cosmic frontier" has now been officially recognized as a key component of particle physics research for OHEP, and the Office has accordingly supported laboratory and university based groups to participate in various experiments in this broad field (Fermi GST, VERITAS, Auger, DES, LSST, JDEM). However, the number of new proposals for such experiments is growing rapidly, and there is legitimate concern that these requests will exhaust available resources unless some appropriate boundaries are drawn. Thus, a great deal of attention has lately been focused on appropriately defining the terms "particle astrophysics" and "cosmology", interpreted as subfields of particle physics. OHEP has emphasized that the primary focus of research they support should be on understanding the fundamental forces and constituents in the Universe, not on modeling astrophysical phenomena associated with individual sources. Nevertheless, it is clear such issues are not entirely separable. For example, one cannot confidently utilize Type 1a supernovae as calibratable standard candles for probing dark energy, without taking some account of the current state of our understanding of the physics associated with the supernova explosion itself. Similarly, the indirect detection of dark matter via WIMP annihilation is only feasible if one has a detailed understanding of the astrophysical backgrounds that may be confused with the putative signals of such processes. Defining the scope of research at the "cosmic frontier" that OHEP will support is a complicated and still unresolved issue. In our view, it is important for OHEP to take an enlightened view of the totality of the scientific problems under investigation, not to invoke absolute boundaries that may strongly limit scientific progress.

---

[1] We are referring here to proposals requesting funding to participate in building an experiment or facility. Proposals for telescope time for follow-up that might be required as part of a proposal that was reviewed and funded through the NSF PNA would, of course, continue to be reviewed by astronomers on the Time Allocation Committee for that telescope.

**Funding Two Communities**

With three agencies, NASA, DOE and NSF now providing support for research at the "cosmic frontier", there is a critical need for coherence and coordination between the particle physics and astrophysics communities with regard to funding requests, which will ensure optimal use of the support available from all three agencies. The agency panels HEPAP/P5 and AAAC, and National Academy surveys and committees are empowered to define and develop a balanced program. In our view the panels, surveys and committees are also in an excellent position to help develop a coherent and coordinated approach to funding this balanced program across the two communities. In addition, workshops and retreats on topics of common scientific interest attended by astrophysicists and particle physicists would be a useful means to develop coherence across, and coordination between, the two communities.

**V. Summary and Conclusion**

Motivated by science questions related to the fundamental constituents of the Universe and the laws of physics that govern them, particle physicists are joining astronomers and astrophysicists to exploit existing instruments and develop new and more sensitive observatories for addressing these questions. A significant and growing fraction of the US research effort in both experimental and theoretical particle physics is devoted to tackling questions related to astrophysics and cosmology[xiv]. Collaborations of astronomers and particle physicists are being formed to develop new state-of-the-art instruments. They are successfully tackling organizational issues that have traditionally been handled differently in the two fields (*e.g.*, membership, authorship, access to data). The agencies that fund particle physics (DOE Office of HEP and NSF EPP and PNA programs) have an established history of supporting scientists to develop the infrastructure needed for large projects -- from initial R&D and project planning, to hardware construction, software development, commissioning and operations. We feel that it is important to provide support to scientists who make similar contributions to very large, long-term projects aimed at addressing questions using astronomical techniques[xvii]. We suggest that within the NSF, proposals submitted by particle physicists to pursue research related to cosmology and dark energy be reviewed within PNA, where research that focuses on dark matter and particle astrophysics is already reviewed and supported. We feel that the priorities of traditional PNA reviewers are an appropriate match to the needs of the very large projects being designed and constructed to address questions related to astrophysics, cosmology and dark energy that motivate particle physicists. We understand that the DOE Office of HEP must define the scope of research that is consistent with its mission; however, we emphasize that addressing questions related to understanding the fundamental forces and constituents in the Universe sometimes require more detailed understanding of astrophysical processes than currently exists. We believe that supporting the astrophysics and particle physics communities to fully collaborate on the planning and construction of new projects will lead to broader scientific reach and higher productivity than either community would achieve working alone. We look forward to contributing to these collaborations in addressing important challenges to our understanding of the Universe.